\input epsf
\documentstyle[11pt,aaspp4]{article}

\singlespace

\def\s-1{{\rm\,s^{-1}}}

\def\spose#1{\hbox to 0pt{#1\hss}}
\def\C3H2{{\rm\,\rm C_3H_2}}
\def\NH3{{\rm\,\rm NH_3}}
\def\HOCO+{{\rm\,\rm HOCO^+}}
\def\lta{\mathrel{\spose{\lower 3pt\hbox{$\mathchar"218$}}
     \raise 2.0pt\hbox{$\mathchar"13C$}}}
\def\gta{\mathrel{\spose{\lower 3pt\hbox{$\mathchar"218$}}
     \raise 2.0pt\hbox{$\mathchar"13E$}}}

\begin{document}

\font\twelvei = cmmi10 scaled\magstep1 
       \font\teni = cmmi10 \font\seveni = cmmi7
\font\mbf = cmmib10 scaled\magstep1
       \font\mbfs = cmmib10 \font\mbfss = cmmib10 scaled 833
\font\msybf = cmbsy10 scaled\magstep1
       \font\msybfs = cmbsy10 \font\msybfss = cmbsy10 scaled 833
\textfont1 = \twelvei
       \scriptfont1 = \twelvei \scriptscriptfont1 = \teni
       \def\mit{\fam1 }
\textfont9 = \mbf
       \scriptfont9 = \mbfs \scriptscriptfont9 = \mbfss
       \def\bmit{\fam9 }
\textfont10 = \msybf
       \scriptfont10 = \msybfs \scriptscriptfont10 = \msybfss
       \def\bmsy{\fam10 }

\def\eg{{\it e.g.}}
\def\ie{{\it i.e.}}
\def\lsim{\raise0.3ex\hbox{$<$}\kern-0.75em{\lower0.65ex\hbox{$\sim$}}} 
\def\gsim{\raise0.3ex\hbox{$>$}\kern-0.75em{\lower0.65ex\hbox{$\sim$}}} 

\def\apj{{ApJ}}
\def\mnras{{MNRAS}}
\def\pasj{{PASJ}}
\title
{Evolution of Tidally Truncated Globular Clusters with Anisotropy}

\author 
{Koji Takahashi$^1$, Hyung Mok Lee$^2$ and Shogo Inagaki$^3$\\
$^1$Department of Earth and Space Science, Graduate School of
Science, Osaka University, Osaka 560,
Japan\\
E-mail: takahasi@vega.ess.sci.osaka-u.ac.jp\\
$^2$ Department of Earth Sciences, Pusan  National  University,
Pusan 609-735, Korea\\
E-mail: hmlee@uju.es.pusan.ac.kr\\
$^3$ Department of Astronomy, Kyoto University, Kyoto 606-01, Japan\\
E-mail: inagaki@kusastro.kyoto-u.ac.jp}

\date{Received 1997; in original form 1997 January }
%
\begin{abstract}
The evolution of tidally truncated globular clusters is investigated by 
integrating two-dimensional Fokker-Planck equation that allows the development 
of velocity anisotropy. We start from the isotropic Plummer model with tidal
cut off and follow the evolution through the core collapse. The heating
by three-body binary is included to obtain the evolution past the core collapse.
The radial anisotropy in velocity dispersion develops during the pre-collapse
evolution in the outer parts of the cluster. 
However, the anisotropy becomes highly depressed during the 
post-collapse evolution because of rapid loss of radial orbits. Maximum
radial anisotropy appears just after the beginning of the expansion, and
degree of anisotropy decreases slowly as the total mass of the cluster 
decreases. The density profiles of pre-collapse and early phase of
post-collapse clusters require King-Michie models while the late phase
of clusters can be well represented by isotropic King models. However,
the detailed radial behavior of degree of anisotropy of our Fokker-Planck
model is somewhat different from that of best-fitting King-Michie models.
\end{abstract}



\section{INTRODUCTION}

Radial velocity anisotropy develops rapidly as a 
result of two-body relaxation in stellar systems (e.g., Spitzer 1987). 
It has been well known that highly anisotropic clusters would show $r^{-3.5}$ 
density profile in the outer parts if the cluster is {\it isolated} 
(see, e.g. Binney \& Tremaine 1987, p522).
The existence of tidal truncation makes the surface density profile 
more complicated than the simple power-laws. 
In order to determine the velocity anisotropy, one needs to measure both 
tangential and radial velocities of large number of stars in a cluster.
Detailed studies of globular clusters reveal that some clusters require
significant degree of radial anisotropy (e.g., Lupton, Gunn \& Griffin 1985)
but many can be fitted by simple
King models which have isotropic velocity distribution throughout the
clusters (e.g., King 1985).

The dynamical evolution of globular clusters has been investigated intensively
in recent years with various numerical techniques. 
Recent studies on dynamical evolution of globular clusters have focused
on the physical processes in the central parts of the clusters where the
anisotropy is rather unimportant. The most popular tool of the study of 
cluster evolution has been the numerical integration of Fokker-Planck equation.
Although Cohn (1979) has pioneered the integration of two-dimensional
Fokker-Planck equation 
in energy ($E$) and angular momentum ($J$) space, difficulties in obtaining
accurate results prevented further use of the two-dimensional Fokker-Planck
equation.  Instead, one-dimensional isotropic Fokker-Planck equation has been 
widely used because it is much easier and faster to integrate.

Most of these difficulties have been lifted by adopting more accurate schemes
for the integration of two-dimensional Fokker-Planck equation (e.g., 
Takahashi 1995). These new schemes have been successfully applied to 
pre-collapse (Takahashi 1995) and post-collapse evolution (Takahashi 1996)
of isolated globular clusters.
As expected from simple theory as well as gas-dynamical models (e.g., 
Spurzem 1996), radial
anisotropy becomes significant as the cluster experiences the dynamical 
evolution.

A natural extension of the two-dimensional Fokker-Planck technique is to 
study the evolution of tidally limited clusters. Here we report the results of
such simulations for single component models.  In the immediately following 
section, we describe our models in detail. The results of our numerical
integrations are presented in \S 3. The final section summarizes our findings.

\section{Description of Models}

We assume only a single mass component in order to maintain the simplicity
for the models. The tidal field experienced by the cluster depends on the
location of the cluster in the Galaxy. However, we assume that the tidal
field is fixed. Thus the cluster is assumed to move along a circular orbit
in the spherically symmetric potential. This is clearly unrealistic because
the cluster orbit is neither circular nor the Galactic potential is spherically
symmetric. The time dependent tidal boundary will affect the 
rate of evaporation, but qualitative results will be somewhat insensitive to
this simplifying assumption. The disk or bulge of the Galaxy 
could also cause gravitational shock which is very efficient
in dissolving the globular clusters (e.g., Gnedin \& Ostriker 1997). 
The effects of tidal shock will be considered
in the forthcoming papers.

The tidal boundary is treated as follows. For a star having energy $E$ and
angular momentum $J$, we can determine the apocenter $r_a$ via

\begin{equation}
J^2 = 2r_a^2 \left[ \phi (r_a) - E \right],
\end{equation}
where $\phi (r)$ is the gravitational potential. 
(In this paper we adopt the sign convention for the potential such that
$\phi>0$ inside the cluster.)
We have assumed that the
star will be lost if the $r_a$ is greater than the tidal radius $r_t$.
The tidal radius is determined by the following condition (e.g., Spitzer
1987),

\begin{equation}
r_t = R_G \left( {3M_C \over M_G}\right)^{1/3},
\end{equation}
where $R_G$ is the Galactocentric radius of the orbit, and $M_C$ and $M_G$
are the masses of the cluster and the Galaxy (within radius $R_G$), 
respectively.

The removal of stars whose $r_a>r_t$ effectively removes those in radial
orbits. We have shown this criteria in $(E,J)$ space for the case of $\phi (r)$
equals to that of Plummer model in Figure 1. Clearly, the stars with 
low $J$ are preferentially removed and the velocity distribution becomes 
isotropic. We call this criteria for the evaporation as `apocenter condition'.

In order to test the sensitivity of the results on the assumption of tidal 
evaporation, we also made a separate assumption for the evaporation based
on energy constraint:
i.e., stars with $E_t= {GM\over r_t} > E$ are assumed to disappear from
the cluster. This is essentially the same assumption for the isotropic
models considered by Lee \& Ostriker (1987) and others.  This energy criterion
is also shown in Figure 1 as dotted lines. Clearly the difference between these
two conditions is somewhat small. For $J=0$, the apocenter condition becomes
identical to the energy condition. 

We have included the heating effects by binaries formed via three-body
processes. The heating effect is included as an additional term in the 
Fokker-Planck equation like other calculations (e.g., Cohn 1985; Lee, Fahlman
\& Richer 1991).
The initial cluster is assumed to be the Plummer model with a tidal
truncation. Initial tidal radius was set to 32 in Plummer model units.
The number of stars in the initial cluster was set to $N=$20000.
This is clearly much smaller than the number of stars in typical
clusters. However, the general evolution of the cluster is nearly
independent of $N$ except for the central density and core radius.
We have chosen such a small $N$ in order to save the computing time: for
clusters with larger $N$ the core collapse proceeds to very high central
density during which the rest of the cluster remains nearly static.
We have employed the
direct integration method of Fokker-Planck equation developed by
Takahashi (1995).

\section{Results}

As in the case of isotropic models, the cluster loses only a small fraction 
of mass until the core collapse. The evaporation becomes important as
the cluster begins to expand as a result of binary heating.  We now discuss
the numerical results of our simulations. For the time being, we will focus
on the results obtained using apocenter condition. For the comparison purposes,
we also have computed the isotropic models based essentially on Lee,
Fahlman \& Richer (1991).

\subsection{Comparison with Isotropic Models}

The evolution of total mass, core, half-mass and tidal radii, and dimensionless 
evaporation rate $\xi$  as a function of time is shown in Figure 2. 
The dimensionless evaporation rate $\xi$ is defined as
\begin{equation}
\xi\equiv -{t_{rh}\over M} {dM\over dt}.
\end{equation}
The time is measured in units of the initial half-mass relaxation time
$t_{rh,i}$, and the radii is in Plummer model units. 
The
result for the isotropic model is indicated as a dotted line.
The collapse times are 17.5~$t_{rh,i}$ and 15.6~$t_{rh,i}$ for
anisotropic and isotropic models, respectively. The longer duration of 
core collapse in anisotropic model is probably due to smaller
rate of collapse ($t_{rc} d\ln \rho_c /dt$, where $t_{rc}$ is the core
relaxation time) for anisotropic models (Louis 1990,
Takahashi 1995).
The cluster loses mass faster in anisotropic model just after the beginning
of the post-collapse expansion. This is due to the fact that the
anisotropy builds up at the time of core-collapse. 
Since the the cluster
just after the core collapse has substantial degree of anisotropy, $r_h$
for a similar $r_t$ (or $M$) cluster is somewhat smaller for anisotropic model 
than for isotropic model. 
While the values of $\xi$ are similar for isotropic and
anisotropic models, the behavior is somewhat different. In anisotropic
model, $\xi$ increases more or less exponentially while the variation is
rather mild for isotropic model. The changing degree of anisotropy has clearly
played some role here, mainly because of the difference in $t_{rh}$.
In general, we conclude that the global behavior of the total mass and
dimensionless evaporation rates are similar between isotropic and anisotropic
models.


Ambartsumian (1938) and Spitzer (1940) estimated the evaporation rate
for an isolated cluster
by assuming that a Maxwellian distribution is established during 
the relaxation time $t_{rh}$.
The rate $\xi$ is set equal to the fraction of stars that have speeds
exceeding the root mean square escape speed 
$\langle v_e^2 \rangle^{1/2} = 2v_m$, 
where $v_m$ is the root mean square speed.
The Ambartsumian-Spitzer (AS) rate may be modified 
for a tidally truncated cluster (Spitzer 1987, p57).
Using the virial theorem,
\begin{equation}
Mv_m^2 = W \equiv k \frac{GM^2}{r_h} ,
\end{equation}
where $W$ is the total gravitational energy and $k \approx 0.4$ for a wide
range of density distribution.
The mean square escape speed $\langle v_e^2 \rangle$ 
in the tidally truncated cluster may be expressed as
\begin{equation}
\langle v_e^2 \rangle = 4v_m^2(1-\gamma) ,
\end{equation}
where
\begin{equation}
\gamma = \frac{1}{2k}\frac{r_h}{r_t} .
\end{equation}
Thus, the modified AS rate is a function of the ratio of the half-mass radius
to the tidal radius, and given by
\begin{equation}
\xi_{\rm AS} = 
\frac{2}{\sqrt{\pi}} x \exp(-x^2) + {\rm erfc} (x) , \label{eq:xi}
\end{equation}
where
\begin{equation}
x = [6(1-\gamma)]^{1/2} , \label{eq:x}
\end{equation}
and ${\rm erfc}$ is the complementary error function.

Figure 3a shows the evolution of $\xi$ again along with $\xi_{\rm AS}$
for the anisotropic model.
In calculating $\xi_{\rm AS}$ we simply set $k=0.5$,
since $k$ is around 0.5 during the post-collapse phase.
Note that the results are insensitive to the choice of $k$.
Figure 3b shows the evolution of the ratio $r_h/r_t$ for this model.
The behavior of $\xi$ after the core collapse is rather well approximated 
by $\xi_{\rm AS}$, i.e., $\xi_{\rm AS}$ increases roughly exponentially.
Figures 3a and 3b show that $\xi$ and $\xi_{\rm AS}$ differ by a nearly 
constant factor of about 2 during the post-collapse phase. 


We note here that similar difference between the AS rate and numerical results
exists in isotropic clusters.
H\'enon (1961) found that $\xi =4.5 \times 10^{-2}$ for his
self-similar model for the post-collapse cluster, while $\xi_{\rm AS}$ is
estimated to be $2.0\times 10^{-2}$ for his model ($r_h/r_t=0.145$).
The reason for having higher values
of $\xi$ than $\xi_{\rm AS}$ may be the expansion of the cluster: the stars
near the tidal boundary becomes unbound to the cluster as a result of
expansion of the cluster.
The evaporation rate in an isolated cluster depends somewhat sensitively
on the distribution function (e.g., Giersz \& Heggie 1994). However, the
evaporation rate of the tidally limited cluster is larger than isolated 
cluster by a large factor. 
Therefore, we may write $\xi = C \xi_{\rm AS}$,
where $C$ is a numerical factor which may depend on the structure of the
cluster and the presence of energy input. 

The evaporation rate depends somewhat sensitively on velocity distribution
for isolated clusters (e.g., Giersz \& Heggie 1994). In our case, we 
find that $\xi$ for anisotropic cluster varies more significantly than
isotropic case. Such a variation can be partially explained by the
changes in cluster structure due to the anisotropy. 
The ratio $r_h/r_t$ increases roughly linearly after the core collapse as 
shown in Figure 3b.
In the right side of equation (\ref{eq:xi}),
the first term is dominant for the adequate range of $x$ for the 
present problem.
(If we take $k=0.5$ and $r_h/r_t=0.1$,
the values of the first and second terms are about 0.012 and 
0.0010, respectively.)
Therefore, $\xi_{\rm AS}$ increases roughly exponentially with 
$r_h/r_t \propto t$, 
unless $C$ changes significantly during the post-collapse evolution.
Figures 3c and d are the same as Figures 3a and b, respectively, but for the 
isotropic model.
Also in the isotropic case, the behavior of $\xi_{\rm AS}$ is not far from 
that of real $\xi$ at the late post-collapse phase.
The variation of the ratio $r_h/r_t$ is mild, 
and the ratio is, roughly speaking, constant throughout almost all of 
the post-collapse phase.
In other words, the post-collapse evolution of the isotropic model is more 
self-similar than that of the anisotropic model.

The different behavior of $\xi$ between the anisotropic and isotropic models 
is thus explained, at least in part, 
by the difference in the evolution of $r_h/r_t$. Obviously the evolution
of the degree of anisotropy has played some role here in making changes to the
structure of the cluster. 
It is still not clear, however, why the ratio increases roughly linearly
with time in the anisotropic model.

\subsection{Degree of Anisotropy}
As we have seen in \S 2, the degree of anisotropy would be reduced
by the presence of the Galactic tidal field because the stars on radial orbits
will be preferentially removed from the cluster. This is clearly demonstrated
in Figure 4 where the degree of anisotropy as a function of radius is shown at
several different epochs. The degree of anisotropy $\beta$ in this figure
is defined as
\begin{equation}
\beta= 1 - {\sigma_t^2 \over \sigma_r^2}
\end{equation}
where $\sigma_t$ and $\sigma_r$ are tangential and radial velocity dispersions,
respectively. The degree of anisotropy generally increases during the 
pre-collapse phase, but it decreases as the cluster loses the mass during the
post-collapse expansion. Note that the re-expansion starts when the cluster mass
becomes about 89\% of the initial mass. That epoch is indicated as a
solid line in this figure. 

The degree of anisotropy increases outward, but reaches maximum at around
half of the tidal radius. The peak value of $\beta$ becomes around 0.6 but
decreases rapidly as the cluster loses mass. Just inside the tidal radius,
$\beta$ becomes negative. This is mainly due to the fact that there should be
turning points near the tidal radius (e.g., Oh \& Lin 1992).
Thus the orbit should be predominantly circular near the tidal boundary. 

Takahashi (1996) noticed that $\beta$ reaches around 0.1 for Lagrangian radius
of 0.2 $M_i$ during the post-collapse phase. Our models with tidal boundary
do not show such an asymptotic behavior. The degree of anisotropy clearly
gets suppressed by the presence of stellar evaporation.

The degree of anisotropy for the models with {\it energy condition} are shown
in Figure 5. The general behavior is similar to the models with {\it apocenter
condition} except for the outer parts. Instead of having monotonically
decreasing behavior after the peak value, $\beta$ reaches minimum and increases
backward.  
Since energy, or isotropic tidal boundary condition is imposed,
$\beta$ should reach zero at the tidal radius.
However, such behavior would not be easily detectable because
of extremely low stellar density near the tidal boundary.

The development of negative-$\beta$ anisotropy at the late evolutionary phase
is in part caused by the change of the gravitational potential.
(Note that $\beta$ becomes negative also under the energy condition 
which does not preferentially remove radial orbits.)
The effect of relaxation is relatively weak at this phase because of 
the low density.
The rate of the energy change of a star of energy $E$ and angular momentum $J$,
due to the adiabatic potential change, is given by
\begin{equation}
\frac{dE}{dt} = \left. \int_{r_p}^{r_a} \frac{dr}{v_r} 
\frac{\partial \phi(r,t)}{\partial t}
\right/ \int_{r_p}^{r_a} \frac{dr}{v_r} , \label{eq:ec}
\end{equation}
where $v_r=[2(\phi-E)-J^2/r^2]^{1/2}$.
At the late stage of the evolution, 
$\partial \phi / \partial t$ is negative and increases with $r$
(note $\phi > 0$).
Then $dE/dt$ is also negative, and, for halo stars,
decreases with $J$ for fixed $E$.
For a halo star, 
the contribution from the region of $r \approx r_a$ is dominant
in the integration in the numerator of equation (\ref{eq:ec}), 
because the star stays for longer time at this region.
Therefore, $dE/dt$ is more negative for higher $J$, or smaller $r_a$.
Thus energy $E$ of circular-orbit stars decreases more 
than that of radial-orbit stars does,
and the tangential velocity dispersion exceeds the radial one in the halo.


\subsection{Structure of Post-collapse Clusters}

Substantial fraction of observed globular clusters have surface brightness
distribution close to that of King models. Although the surface brightness
distribution is not very sensitive to the
velocity anisotropy,
highly anisotropic models cannot be fitted to isotropic King models. Here
we discuss the structure of post-collapse models with apocenter condition
for stellar ejection since the structure of the cluster is not sensitive
to the choice of criteria for ejection.

The degree of anisotropy is greatest just before the post-collapse
expansion an shown in Figure 4. The total cluster mass in this phase
is about 89\% of the initial mass.
Isotropic King model gives asymptotic slope of -2 in log$\rho$ versus
log$r$ plot for highly concentrated models, but highly anisotropic 
pre-collapse cluster has steeper slope. For example, the density profile of
the Figure 6 has the slope of -2.45. Thus it is difficult
to fit the pre-collapse cluster to an isotropic King model.
Instead, we have fitted the density profile of the pre-collapse
cluster to a King-Michie model (Fig. 6) with  $W_0=15$ and $r_{an}/r_c =300$,
where $r_{an}$ is the anisotropic radius. 
Obviously, King-Michie model gives much better fit than the isotropic King
model.

However, the radial profile of
degree of anisotropy of our Fokker-Planck model does not match well 
with that of best-fitting King-Michie model,
as shown in Figure 7. The
main difference is the presence of plateau of $\beta$ outside the core radius
for the Fokker-Planck result. The
penetration of anisotropy to such a small radius was noticed by Louis \& 
Spurzem (1991) and Giersz \& Spurzem (1994) using gaseous models and by
Takahashi (1995) using Fokker-Planck model.
As the core collapse proceeds, 
the central density-cusp accompanied with constant anisotropy $\beta=0.08$ 
extends into the inner region as shown in the above mentioned works.
The King-Michie model does not give any such behavior in radial anisotropy
profile. The degree of anisotropy of the King-Michie model beyond $r_{an}$
is greater than the Fokker-Planck result. Although the
King-Michie models are  a convenient representation of stellar systems
with radial anisotropy, the runs of anisotropy are significantly
different from those produced by dynamical relaxation.

As evident from Figures 4 and 5, the degree of anisotropy decreases
as the cluster loses mass. In Figure 8, we have shown the density profiles
at $M/M_i=0.7$ and 0.5. Also shown here are density profiles of isotropic
King models having same concentration (dotted lines) as well as best fitting 
King-Michie models (broken lines). Although the isotropic King models give 
reasonable fit to our numerical results, 
the halo is better fitted when small anisotropy is introduced. In fitting
to King-Michie models, $r_{an}/r_c$ is not very precisely constrained 
because the deviation from isotropic King model is rather small. 

As the cluster loses significant mass, the clusters become more isotropic.
In Figure 9, we have compared the density profiles at two different
epochs ($M/M_i =0.3$ and 0.12) with the isotropic King models.
As in Figure 8, the King models are chosen to have the same concentration 
parameter $c\equiv \log(r_t/r_c)$.
It is clear that the King models can represent the late phase of 
post-collapse clusters
very well. This is simply because the degree of anisotropy is very small
during the post-collapse phase. Thus we may conclude that the clusters with
less than than 50\% of the initial mass would have negligible degree of
anisotropy.

\section{Summary}
We have carried out anisotropic Fokker-Planck calculations for the dynamical 
evolution of globular clusters in static Galactic tidal field. 
We have considered simple model of single mass component and included
the effect of heating by binaries formed by three-body processes in order
to obtain the post-collapse evolution. The initial cluster was assumed to be
Plummer model with a tidal truncation. As in the case of 
isotropic cluster models with tidal boundary, the cluster mass decreases
roughly linearly in time. Since evaporation time is much larger than the
time to core collapse, most mass loss occurs during the
post-collapse phase. We have assumed that the stars with
apocenter distance greater than the tidal boundary are ejected from the
cluster. 
This `apocenter condition' for the stellar ejection provides preferential
removal of stars in radial orbits. We have followed the cluster evolution 
until the cluster loses about 99\% of the initial mass.

The radial anisotropy develops as a result of two-body relaxation in the
outer parts. However,
we found that the velocity anisotropy in the halo is usually smaller
than the isolated models by the presence of Galactic tidal field.
During the post-collapse expansion, the degree of anisotropy
decreases monotonically with time. The density profiles of
clusters in the late phase of post-collapse evolution are 
found to be well approximated by the
isotropic King models because the radial anisotropy is not
significant.

In order to find the sensitivity of the results on the assumption of
tidal evaporation process, we have also applied different criteria for the
stellar ejection: in an alternative model we assumed that the stars are 
removed from the
cluster if $E< E_{tid}=GM/r_t$. This is essentially the same assumption
used in isotropic models considered by Lee \& Ostriker (1987). This
criteria is referred as `energy condition'.

The results of `energy condition run' are compared with the apocenter 
constraint run. The 
general behavior of the cluster evolution is not much different between 
these two different models except for outer parts near the
tidal boundary. The degree of anisotropy is similarly depressed
due to the presence of tidal boundary although the energy condition
does not preferably removes the stars on radial orbits.
This is found to be 
due to the fact that most weakly bound stars susceptible for ejection
are on radial orbits.
Thus we conclude that the suppression of degree of anisotropy is not mere
a consequence of the `apocenter' constraint.

Considering the facts that the cluster mass decreases
linearly and the degree of anisotropy decreases in time
during the post-collapse phase, we may judge the evolutionary
status of the cluster by measuring the degree of velocity 
anisotropy. 
However, we need to extend our models to more realistic models in order
to reach more quantitative conclusion. We are planning to study 
models including tidal shock by disk and bulge and
initial mass function. 

\vskip 0.2truein
KT was supported by the Grant-in-Aid for Encouragement of Young
Scientists by the Ministry of Education, Science, Sports and Culture of Japan 
(No. 1338).
HML was supported by the KOSEF Grant through Grant No. 95-0702-01-01-3. 



\clearpage
\begin{figure}
\plotone{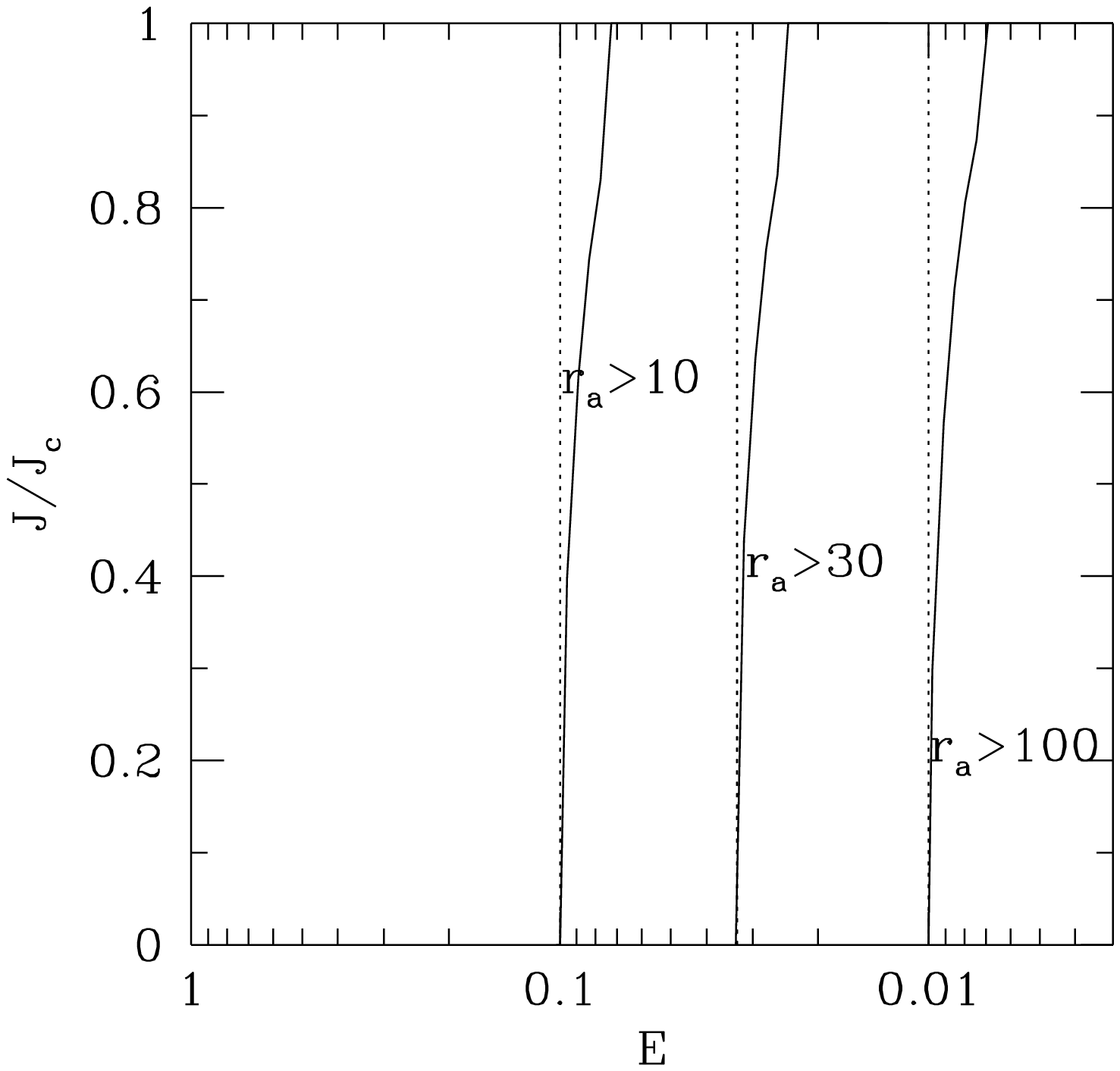}
\figurenum{1}
\caption{The criteria for evaporation of stars in $(E,J)$ space for
$r_{tid}=$10, 30 and 100. The underlying potential is assumed to be that of
a Plummer model and the radius is in Plummer model units. The solid line
represents the {\it apocenter condition} while the dotted line represents
the {\it energy condition}.}
\end{figure}

\begin{figure}
\plotone{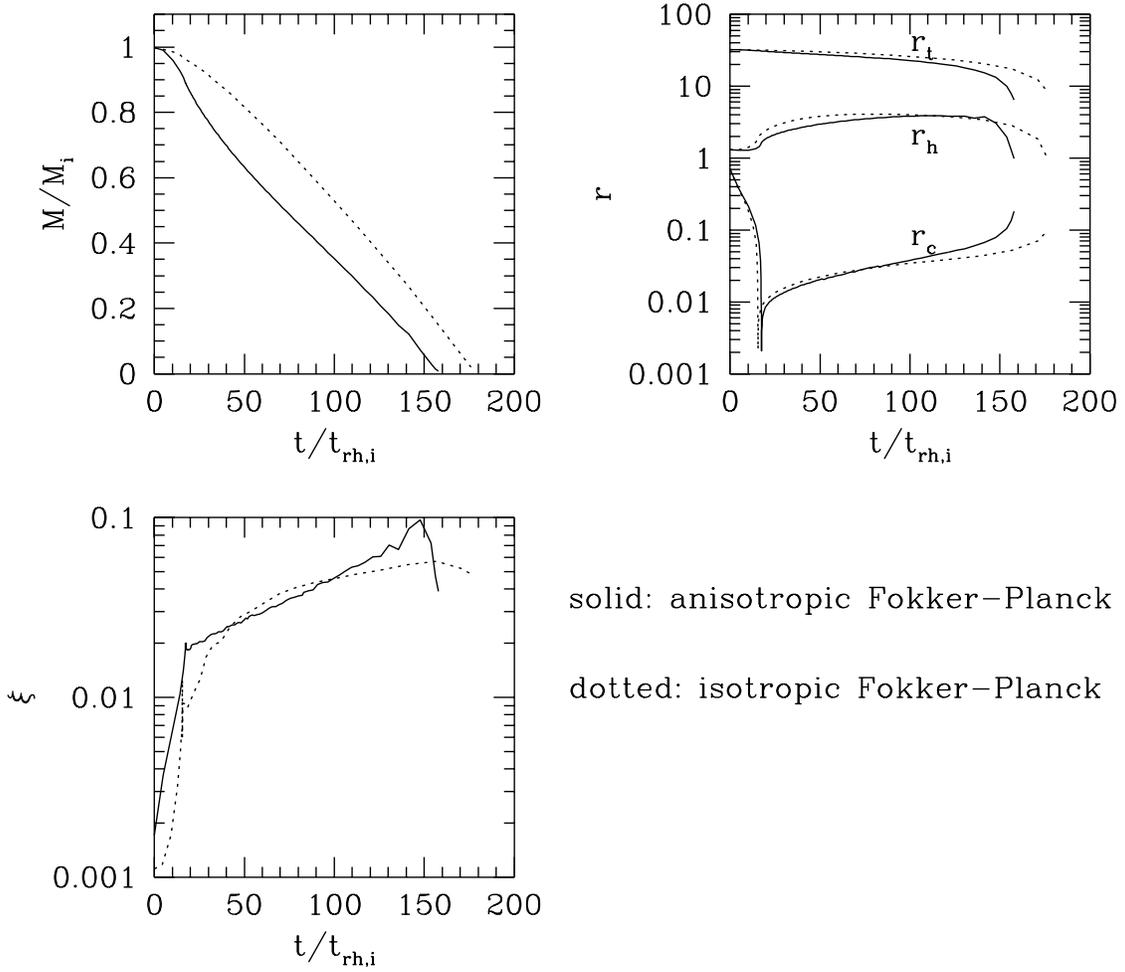}
\figurenum{2}
\caption{
Evolution of the (a) total mass,  (b) core, half-mass and tidal
radii, and (c) $\xi$ as a function of time. The solid line
is for the anisotropic result and the dotted line is for the isotropic
result.
}
\end{figure}

\begin{figure}
\plotone{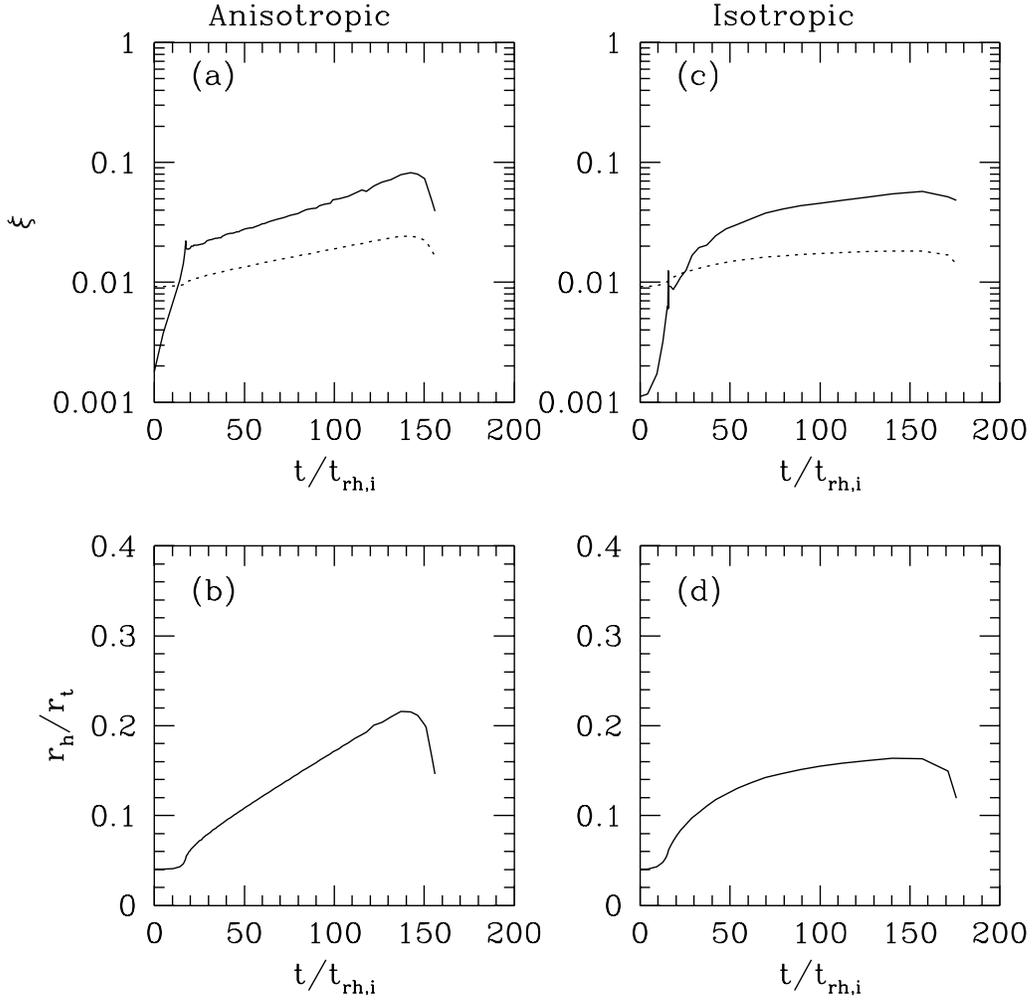}
\figurenum{3}
\caption{Evolution of (a) $\xi$ (solid) and $\xi_{\rm AS}$ (dotted), 
and (b) $r_h/r_t$
for the apocenter condition anisotropic model. 
(c), (d) The same as (a) and (b), but for the isotropic model.}
\end{figure}

\begin{figure}
\plotone{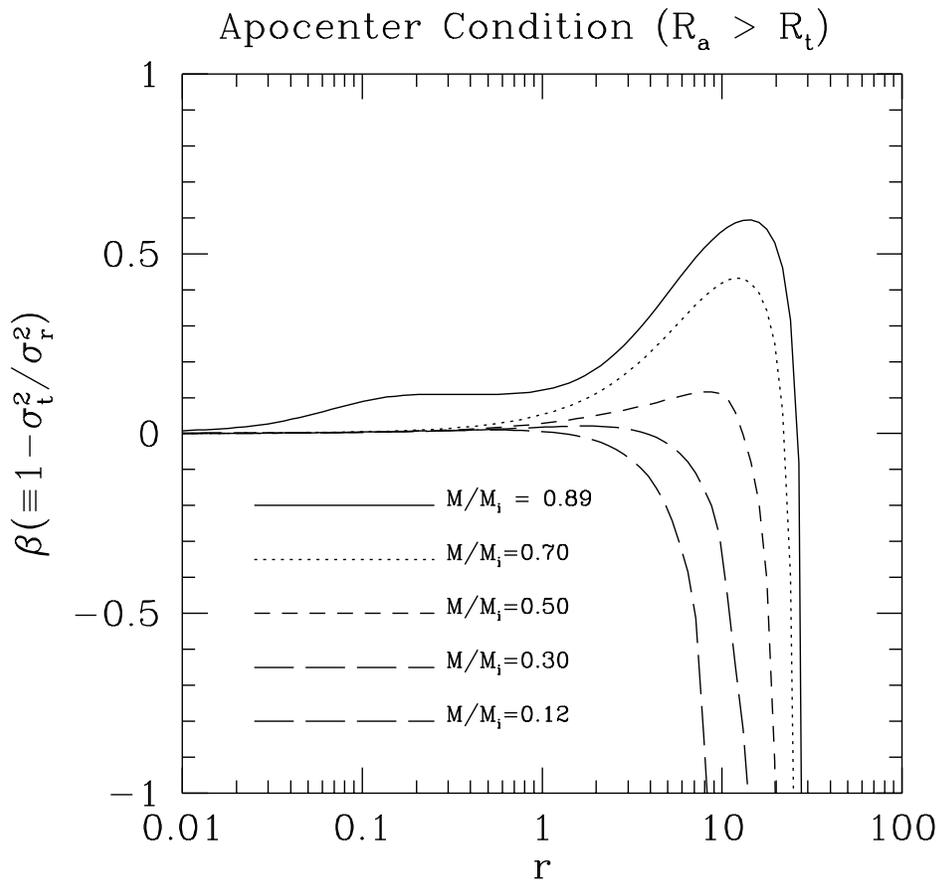}
\figurenum{4}
\caption{The degree of anisotropy as a function of radius for several
different epochs for apocenter condition.}
\end{figure}

\begin{figure}
\plotone{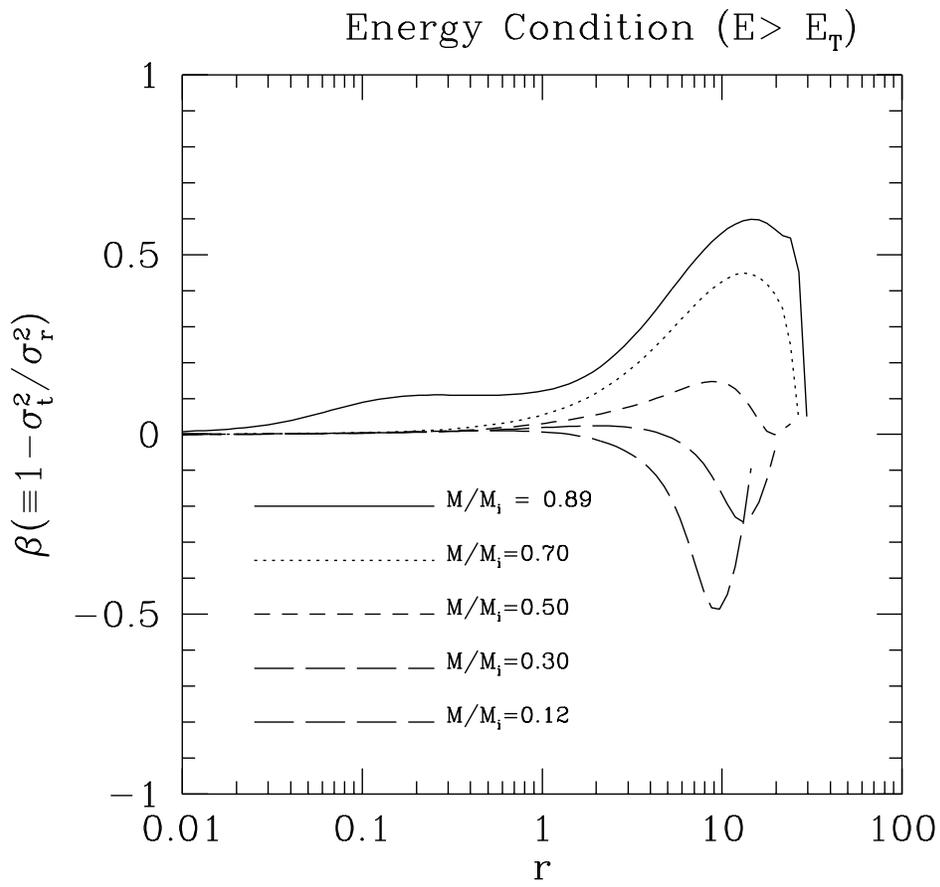}
\figurenum{5}
\caption{
Same as Figure 4 except that the energy condition is used for
evaporating stars.
}
\end{figure}

\begin{figure}
\plotone{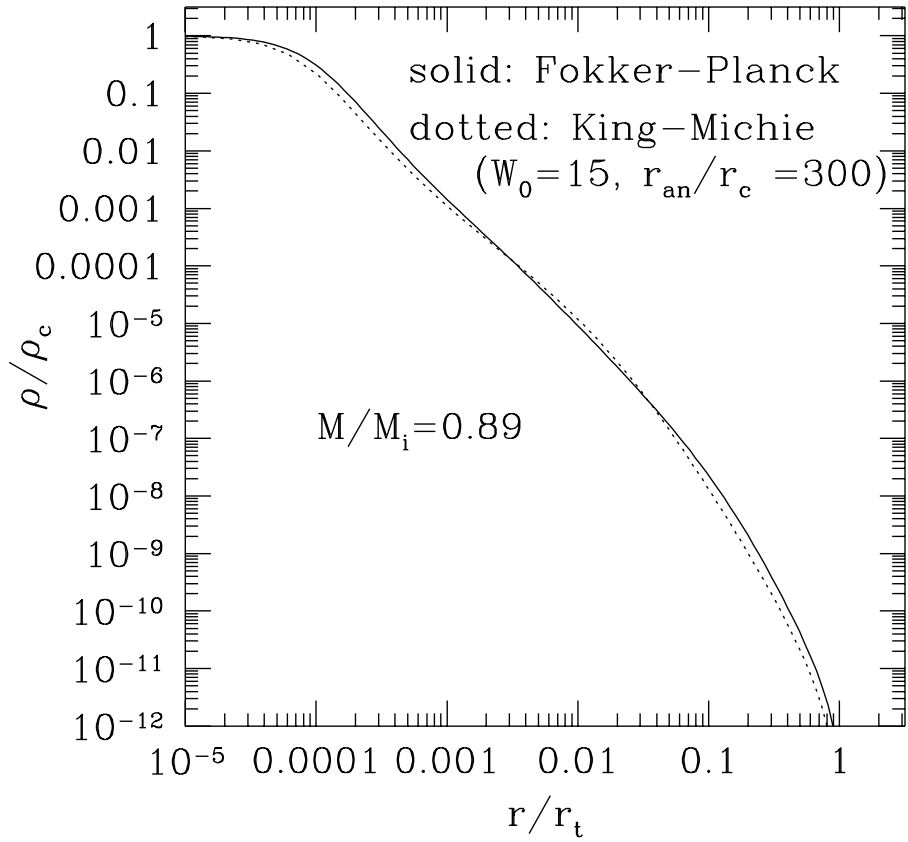}
\figurenum{6}
\caption{ The density profile at near the peak core-collapse compared 
with the best fitting King-Michie model.}
\end{figure}

\begin{figure}
\plotone{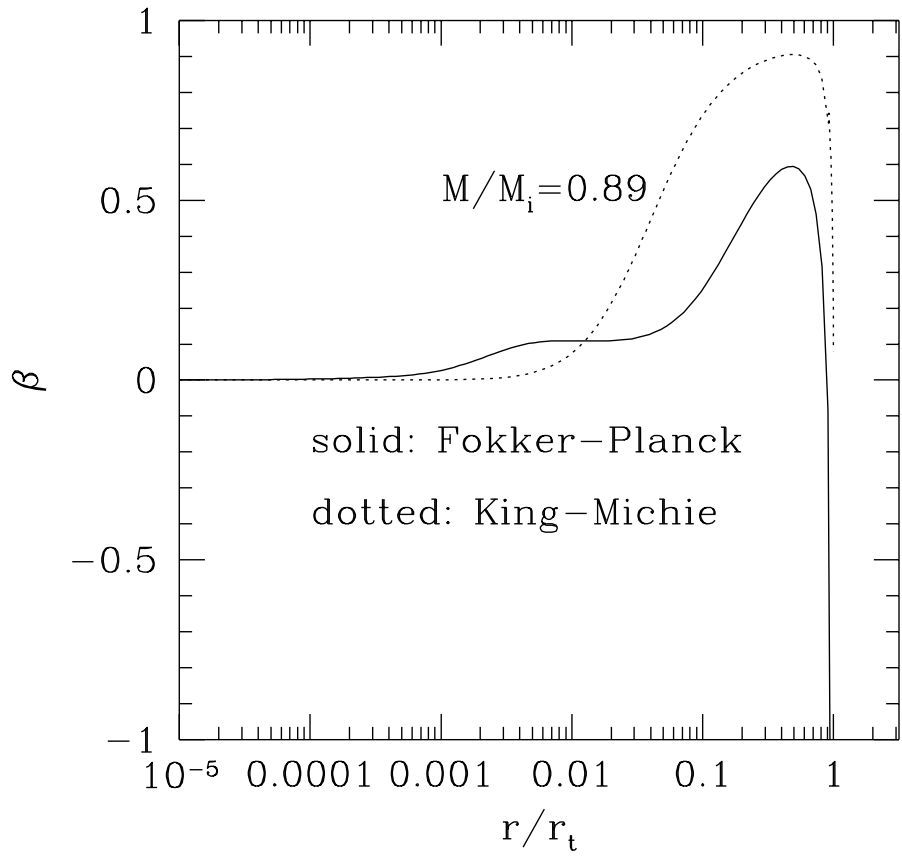}
\figurenum{7}
\caption{Radial distribution of degree of anisotropy for collapsed cluster (solid)
and for best-fitting King-Michie model (dotted).
}
\end{figure}

\begin{figure}
\plotone{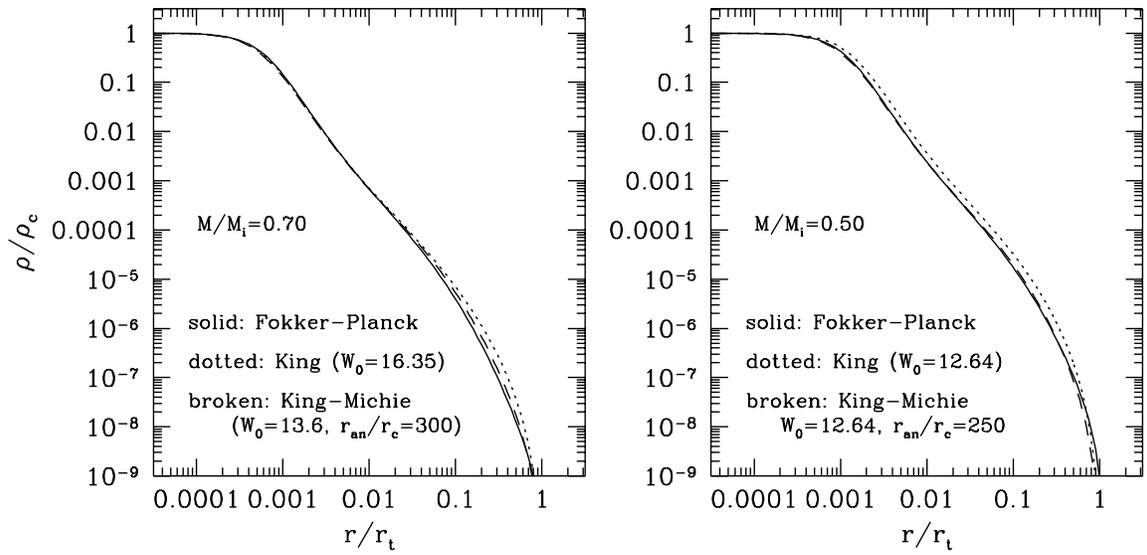}
\figurenum{8}
\caption{  The density profiles at two early epochs of post-collapse phase
compared
with the isotropic King models with the same degree of concentration (dotted
lines) and best-fitting King-Michie models (broken lines).}
\end{figure}

\begin{figure}
\plotone{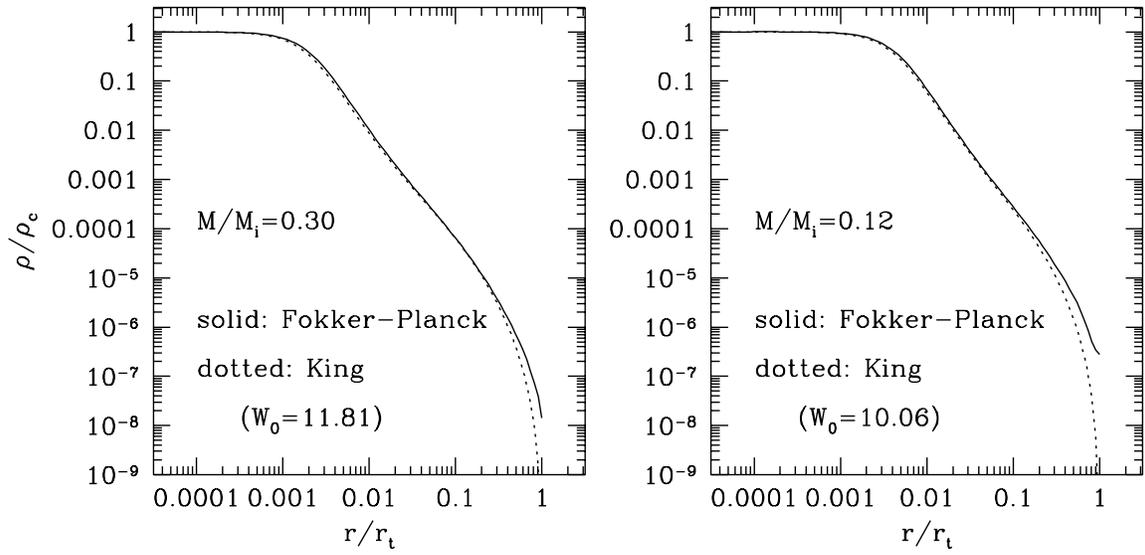}
\figurenum{9}
\caption{The density profiles at two late epochs of post-collapse phase
compared
with the isotropic King models with the same degree of concentration (dotted
lines).}
\end{figure}

\end{document}